\documentclass[usenatbib,usegraphicx]{mn2e}
\usepackage{amsmath}
\usepackage{hyperref}
\title[Radio and X-ray from A2319]{A Radio and X-ray Study of the Merging Cluster A2319}

\author[Storm, Jeltema, \& Rudnick]
{
Emma Storm$^1$, Tesla E. Jeltema$^{1,2}$, Lawrence Rudnick$^3$\\
$^1$Department of Physics,  University of California, 1156 High St., Santa Cruz, CA 95064, USA\\
$^2$Santa Cruz Institute for Particle Physics,  University of California, 1156 High St., Santa Cruz, CA 95064, USA\\
$^3$Minnesota Institute for Astrophysics, School of Physics and Astronomy, University of Minnesota,\\116 Church Street SE, Minneapolis, MN 55455, USA
}

\begin{document}
\maketitle



\begin{abstract}
A2319 is a massive, merging galaxy cluster with a previously detected radio halo that roughly follows the X-ray emitting gas. We present the results from recent observations of A2319 at $\sim20$~cm with the Jansky Very Large Array (VLA)  and a re-analysis of the X-ray observations from \textit{XMM-Newton},  to investigate  the interactions between the thermal and nonthermal components of the ICM . We confirm previous reports of an X-ray cold front, and report on the discovery of a distinct core to the radio halo, $\sim800$~kpc in extent, that is strikingly similar in morphology to the X-ray emission, and drops sharply in brightness at the cold front.  We detect additional radio emission trailing off from the core, which blends smoothly into the  $\sim2$~Mpc halo detected with the Green Bank Telescope (GBT) \citep{Farnsworth2013}.  We speculate on the possible mechanisms for such a two-component radio halo, with sloshing playing a dominant role in the core. By directly comparing the X-ray and radio emission, we find that a hadronic origin for the cosmic ray electrons responsible for the radio halo would require a magnetic field and/or cosmic ray proton distribution that increases with radial distance from the cluster center, and is therefore disfavored.
\end{abstract}


\section{Introduction}\label{sec:int}

Galaxy cluster mergers are among the most energetic events in the universe. Major mergers between massive clusters ($\sim10^{15}M_{\odot}$) drive shocks and generate turbulence throughout the intracluster medium (ICM), thus providing potential acceleration sites for relativistic particles. Observations of diffuse synchrotron radiation in the radio on Mpc scales at $\sim1$~GHz demonstrate that clusters host a population of relativistic, GeV electrons and $\mu$G-scale magnetic fields distributed throughout the ICM. \textit{Giant radio halos} fill the cluster volume, tend to trace the ICM of clusters, are unpolarized, and have steep spectra, with $\alpha \ga 1$, where the flux density $S_{\nu} \propto \nu^{-\alpha}$ (for a review of radio emission from clusters, see e.g., \citealt{Feretti2012}). 

Signatures of cluster mergers can be observed in the X-ray emitting ICM. Cluster mergers disturb the ICM gas, which leads to a significant amount of substructure present in the X-ray emission \citep{Schuecker2001,Markevitch2001}. Shocks are one indicator of a dynamically disturbed ICM.  Cold fronts, characterized by a surface brightness discontinuity in the X-ray, occur when a cold subcluster core moves through hotter ambient gas or result from sloshing of the central cool gas in the aftermath of a merger \citep{Markevitch2007}. The temperature structures of merging clusters tend to be complex, with colder gas tracing the paths of the subcluster cores and heated gas perpendicular to the merger axes \citep{Govoni2004a}.

There exists a strong correlation between the existence of radio halos and the dynamical state of clusters derived from X-ray observations (e.g., \citealt{Cassano2013} and references therein). Radio halos are clearly associated with merging clusters, while more relaxed clusters do not host radio halos \citep[e.g.][]{Cassano2010a}, with few exceptions \citep{Bonafede2014}. Radio halos are therefore intimately tied to the dynamical history of clusters, and the origins of radio halos can be effectively probed by studying cluster dynamics, especially with X-ray observations. 

The origin of the cosmic rays responsible for radio halos is still under debate (see \citealt{Brunetti2014} for a review of cosmic ray accleration mechanisms in clusters). In the hadronic model, cosmic ray protons, accelerated by merger-driven shocks and turbulence, fill the volume of the cluster \citep{Volk1996,Berezinsky1997}. These cosmic ray protons collide with thermal particles in the ICM, producing pions that decay to electrons and positrons, which then lose energy \textit{in situ}, via synchrotron radiation if the magnetic fields are sufficiently strong \citep{Dennison1980,Blasi1999}. The hadronic model provides a natural explanation for the diffuse nature of radio halos and for the strong observed correlation between X-ray and radio emission in clusters, since both trace the gas density in this scenario. However, among the products of cosmic ray proton collisions are gamma rays, and clusters have not yet been detected in the gamma-ray band (most recently, \citealt{Ackermann2014}). In order to reproduce the observed synchrotron radio emission in some clusters, the magnetic fields need to be stronger than those inferred from current Faraday Rotation measurements, so that the expected gamma-ray flux does not exceed current upper limits \citep{Jeltema2011,Brunetti2012b}.

In the reacceleration model, a long-lived mildly relativistic population of seed electrons are reaccelerated to energies sufficient to produce observable synchrotron emission by merger-driven turbulence throughout the cluster \citep{Brunetti2001,Brunetti2004,Brunetti2011a,Petrosian2001,Donnert2013}. In this context, the predicted gamma-ray emission from Inverse Compton (IC) scattering is low compared to observed upper limits \citep[e.g.,][]{Brunetti2011a,Brunetti2012b}. However, the properties of turbulence in the ICM are poorly understood, which limits the predictive capabilities of this model.

In this paper we study A2319, a massive, merging, nearby galaxy cluster ($z=0.0557$; \citealt{Truble1999}). Optical observations reveal two subclusters, the more massive A2319A and a smaller subcluster to the northwest, A2319B, separated by $\sim10'$ in the plane of the sky and by $\sim3000$~km~s$^{-1}$ in velocity space \citep{Faber1977,Oegerle1995}. A mass ratio of 3:1 is derived for the A and B subclusters in \citet{Oegerle1995}.

A2319 hosts a previously detected $\sim10'$ ($650$~kpc) radio halo that closely traces the X-ray emission from the ICM \citep{Harris1978,Feretti1997}. However, recent observations with the Green Bank Telescope reveal the true extent of the halo to be $\sim35'$ ($\sim2$~Mpc) across \citep{Farnsworth2013}.

A2319 has been studied extensively in the X-ray by several instruments, including \textit{ASCA} \citep{Markevitch1996}, ROSAT \citep{Feretti1997}, \textit{BeppoSAX} \citep{Molendi1999}, \textit{Chandra} \citep{Govoni2004a,OHara2004}, \textit{Suzaku} \citep{Sugawara2009}, and \textit{XMM-Newton} (\citealt{Ghizzardi2010}; this work). The X-ray emission also reveals several signatures of merger activity, including a complex temperature structure and a cold front to the SE of the central X-ray core of A2319A \citep{Govoni2004a,OHara2004,Ghizzardi2010}. While an optical analysis by \citet{Oegerle1995} claims that the there is a non-negligible chance the subclusters are not actually gravitionally bound, a photometric study of the galaxies in A2319 combined with the detection of a cold front in the X-ray suggests that the cluster is post-merger viewed in projection ($\sim30-70^{\circ}$ to the plane of the sky; \citealt{OHara2004,Yan2014}).

In this paper we present a joint analysis of radio and X-ray observations of A2319. From $\sim20$~cm radio observations with the upgraded Jansky Very Large Array (VLA), we report  more extensive halo emission than previously seen by interferometer measurements, and the discovery of a distinct 800~kpc core to the halo emission and an extension to the southwest. We present a new analysis of archival X-ray observations from \textit{XMM-Newton} to examine potential connections between the radio and X-ray emission in this cluster. We find that the radio halo core traces the central X-ray emission remarkably well, and trails off into a larger-scale region ($2$~Mpc) that corresponds to the emission detected by the GBT. In light of this new discovery of a multi-component halo, we revisit the dynamical history of this cluster and explore possible origin models for this radio halo.

This paper is organized as follows. In Section~\ref{sec:rad}, we review radio observations of A2319 from the literature and present our results from a new analysis of VLA data. In Section~\ref{sec:Xray}, we summarize previous X-ray analyses of A2319 and present a new analysis of archival \textit{XMM-Newton} observations. In Section~\ref{sec:disc}, we discuss the implications of the radio and X-ray observations in the context of cluster dynamics, cosmic ray origins, and magnetic field structure. We conclude in Section~\ref{sec:end}. We adopt a $\Lambda$CDM cosmology, where $H_o = 70$~km~s$^{-1}$~Mpc$^{-1}$, $\Omega_m = 0.3$, $\Omega_{\Lambda} = 0.7$. At the redshift of A2319 ($z=0.0557$), $1''$ corresponds to $1.08$~kpc.

\section{Radio Analysis}\label{sec:rad}

\subsection{Previous Observations}

The radio halo in A2319 has been observed previously with the WSRT and the VLA \citep{Harris1978,Feretti1997}. After subtraction of discrete sources, \citet{Harris1978} reported a $\sim10'$ or $650$~kpc halo with an integrated flux density of $1$~Jy at $610$~MHz using WSRT. Observations at $90$~cm ($330$~MHz) by \citet{Feretti1997} with WSRT and VLA were badly compromised by sidelobes from Cygnus A. The best map was obtained from the WSRT observations at $20$~cm ($1400$~MHz), which yielded a $\sim15'$ or $1000$~kpc radio halo that traced the X-ray emission as observed with ROSAT. The total flux of the halo reported was $153$~mJy after point source subtraction, with an rms noise of $0.035$~mJy beam$^{-1}$ for a $29.0''\times20.4''$ beam. \citet{Feretti1997} noted that they did not capture the full size or flux from the halo due to missing short baselines. \citet{Feretti1997} also reported on a detection of the halo at $408$~MHz with the Northern Cross Radio Telescope (NCRT), which yielded a total halo flux of $1.45$~Jy after point source subtraction.

Observations of the halo in A2319 with the Green Bank Telescope (GBT) were presented in \citet{Farnsworth2013}. The detected halo flux and size were more than double the previous detection with WSRT. \citet{Farnsworth2013} reported a halo flux of $328\pm28$~mJy and a largest angular size of $35'$ (largest linear size of $2$~Mpc) at $1400$~MHz, for a $9.7'\times 9.5'$ beam. Since it is a single dish, the GBT can capture all of flux from extended, diffuse sources such as radio halos. This detection represents the total flux and extent of the halo in A2319. However, the GBT cannot map smaller scale structure in the radio halo because its resolution is poor compared to interferometers.

\subsection{VLA Analysis}

We observed A2319 with the VLA in 2010 in the C and D configurations over two $128$~MHz spectral windows centered on $1348$~MHz and $1860$~MHz. Two pointings were made for each configuration, centered on the subclusters A2319A and A2319B. Pointing centers were $\alpha=19h21m15s.00$, $\delta=43^{\circ}52'00''.00$ and $\alpha=19h20m45s.00$, $\delta=44^{\circ}03'00''.00$. The total time on source was $\sim4.5$~hours for the C configuration and $\sim7$~hours for the D configuration. The data were taken while the new correlator was still being debugged, which resulted in some problems with the analysis, as described below. Data analysis and imaging were performed with the NRAO analysis package CASA\footnote{casa.nrao.edu}, version 4.0.1.

Data from the C configuration were not used as we originally intended. We planned to subtract the fluxes from the point sources in the C configuration images from the D configuration images. However, after calibration and imaging, it was discovered that the fluxes in the C configuration data set were corrupted, and could not be salvaged. We were able to use the C configuration images as guides for locating point sources, in addition to the NVSS (Northern VLA Sky Survey; \citealt{Condon1998}).

D configuration data were calibrated with CASA. 3C286 was used as the flux calibrator and J1845+4007 was used as a bandpass and phase calibrator, which was observed every $20$ minutes. This observation was made in spectral line mode (as are all new VLA observations) with a channel width of $2$~MHz. This allows for more precise excision of radio frequency interference (RFI). The data were Hanning smoothed and RFI was excised first automatically using the \texttt{flagdata} and \texttt{flagcmd} tasks in CASA, and then the remaining RFI was carefully removed by hand. Approximately $50\%$ of the data in each spectral window were contaminated by RFI, which is typical for L band observations. After calibration, the data sets were time-averaged to $10$s from $1$s to speed up image processing.

Imaging was performed using the CLEAN task in CASA. We first created maps using only {\it uv} data at baselines longer than $200\lambda$, to preserve the flux of compact sources while significantly reducing the halo emission. In Table~\ref{tab:vlaps} we list the compact sources located within the $1348$~MHz detected halo region. We scaled their fluxes to $1400$~MHz to facilitate comparison with the NVSS, by first averaging the primary-beam-corrected fluxes from the two pointings and then interpolating between $1348$ and $1860$~MHz. For the sources also found in NVSS, our fluxes agree within calibration uncertainties of a few percent.

\begin{table*}
  \begin{minipage}{6.5in}
    \caption{Table 1: Radio Source Properties}
    \label{tab:vlaps}
    \begin{tabular*}{6.5in}{@{\extracolsep{\fill} } c c c c c c c }\hline
      &                   &    RA             &     Dec           &         VLA             &        &     \\  
      &NVSS ID            &    (J2000)        &     (J2000)       &         (1400 MHz, mJy) & F97    & HM78    \\      \hline
      1  &192004+440034   &    290.01775      &     +44.00958     &         4.0             &  ...   & 252     \\
      2  &192012+435955   &    290.05067      &     +43.99875     &         1.6             &  ...   & 257     \\
      3  &192015+440305   &    290.06508      &     +44.05153     &         87              &  B     & 259     \\
      4  &192017+434851   &    290.07446      &     +43.81422     &         3.9             &  C     & 262     \\
      5  &192053+435232   &    290.22371      &     +43.87572     &         33              &  ...   & 270     \\
      6  &192109+435307   &    290.28854      &     +43.88544     &         25              &  K     & 273     \\ 
      7  &192112+435640   &    290.30217      &     +43.94469     &         27              &  ...   & 277     \\ 
      8  &192118+435817   &    290.32808      &     +43.97156     &         3.5             &  ...   & 278     \\
      9  &192132+435946   &    290.38425      &     +43.99633     &         4.0             &  N     & 282     \\
      10 &192133+435805   &    290.38858      &     +43.96819     &         110             &  ...   & 283     \\
      11 &192142+435749   &    290.42833      &     +43.96375     &         13              &  R     & 291     \\ 
      12 &...             &    290.408        &     +43.9124      &         2.5             &  ...   & 287     \\
      13 &...             &    290.273        &     +44.0798      &         2.5             &  H     & 272     \\
      14 &...             &    290.134        &     +43.9142      &         2.2             &  ...   & 266     \\
      15 &...             &    290.134        &     +43.8942      &         2.0             &  ...   & 265     \\
      16 &...             &    290.115        &     +43.8747      &         1.4             &  ...   & 264     \\ \hline
    \end{tabular*}
    
    \textbf{Notes}
    Column 1: ID number. Column 2: NVSS ID \citep{Condon1998}. Column 3 and 4: Coordinates of radio source; NVSS coordinates given if source is identified in NVSS. Column 5: Source flux measured by VLA D configuration (only baselines longer than 200$\lambda$ present), scaled to $1400$~MHz. Beam is $48''$. Column 6: ID corresponding to the label listed in Table 4 of \citet{Feretti1997}, which only lists radio sources associated with optically-identified cluster member galaxies. Column 7: ID corresponding to the serial number listed in Table 6 of \citet{Harris1978}.
  \end{minipage}
\end{table*}

We then subtracted the baseline-restricted clean components from the full {\it uv} data set, so that we were left with flux only from the halo plus residual noise. We then {\it uv} tapered to a $\sim120''$ beam to enhance the sensitivity to extended emission. We used multiscale CLEAN to image each map. We attempted several iterations of self calibration (phase only) and widefield CLEANing, however these techniques did not noticeably improve image quality, so our final images do not include these processing steps. We mosaiced the CLEANed images from the 2 pointings and applied the primary beam correction.

\begin{figure} 
  \centering
  \includegraphics[scale=0.36]{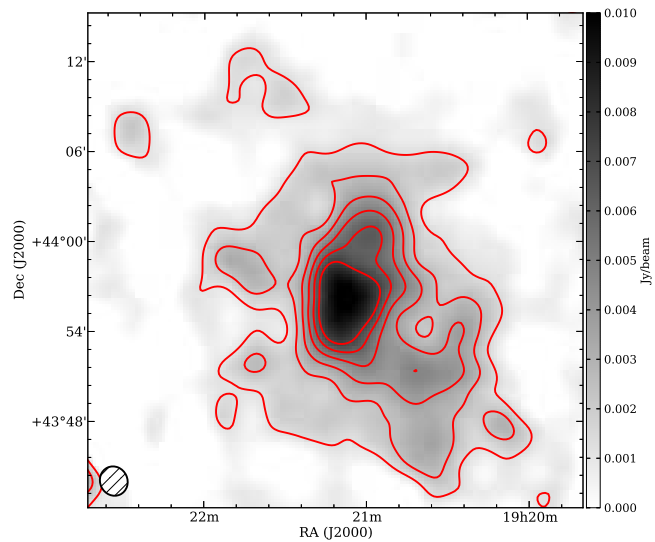}
  \caption{A2319 Halo at 1348 MHz from VLA. Contours in red are (3,6,9...)$\times0.4$~mJy~beam$^{-1}$. Beam is $119'' \times 110''$, shown in black in bottom left.}
  \label{fig:spw0}
\end{figure}

The 1860~MHz spectral window suffered from significant residual RFI which particularly created problems for the reconstruction of the diffuse emission.  Therefore, in the remainder of the paper, we will report only the results from the 1348~MHz map. The resulting image of the diffuse emission at $1348$~MHz is shown in Figure~\ref{fig:spw0}.

\subsection{The Radio Halo in A2319}

The radio halo is significantly larger with a more complex morphology than previously detected in interferometer maps. The flux density within the $3\sigma$ contours  at $1348$~MHz is $240 \pm 10$~mJy, with an rms noise of $0.4$~mJy~beam$^{-1}$. This is significantly less than detected on the GBT by \cite{Farnsworth2013} because of insufficient short {\it uv} spacing data with the VLA. The reported uncertainty in the integrated flux density does not take into account any uncertainties in calibration or imaging. The halo's longest dimension as detected by the VLA at $1348$~MHz is $22'$ or $1400$~kpc, compared to about $35'$ or $2000$~kpc for the GBT. 

\subsubsection{Halo Structure}

\begin{figure}
  \centering
  \includegraphics[scale=0.36]{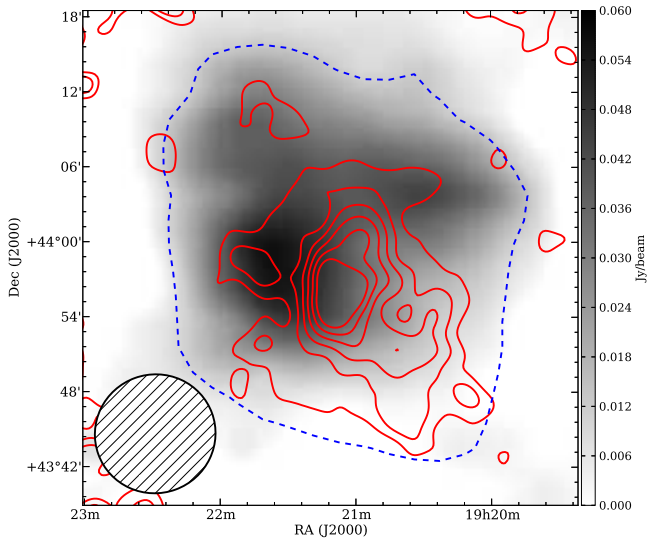}
  \caption{Comparison of VLA (beam: $120''$) and GBT (beam: $570''$) images. The VLA contours are in red, at (3,6,9...)$\times0.4$~mJy~beam$^{-1}$. The lowest contour from the full GBT image ($18$~mJy~beam$^{-1}$) is shown in dashed blue. The grey scale image shows the residual GBT image after subtracting a convolved version of the VLA image. The GBT beam is shown in the bottom left.}
  \label{fig:GBT-VLA}
\end{figure}

Figure~\ref{fig:GBT-VLA} shows the various components of the halo. The full GBT emission spans $2$~Mpc and is shown as a single contour here. The residual GBT emission, after subtracting out the VLA image convolved with the GBT beam, is visible on three sides of the core. On the fourth side, to the southwest, the VLA recovers all the flux seen at the GBT.
This large SW extension was previously undetected by interferometers. With a flux density of $62$~mJy over an area of $\sim 3.6\times 10^{5}$~arcsec$^2$ (about a third of the total area of the halo), it contributes only $25\%$ of the halo flux visible to the VLA. The SW extension appears to have no X-ray counterpart, as discussed in Section~\ref{sec:Xray}.

In this work, we were able to increase the surface brightness sensitivity by convolving down to $120''$ resolution after compact source removal. Previous interferometer images were able to detect the brighter regions of diffuse emission, but were not able to pick out the various sub-structures because of confusion from compact radio emission \citep{Feretti1997}. A hint of the core of the halo may be visible in the \citet{Feretti1997} $90$~cm map, but is likely confused with nearby compact emission (source K, Table~\ref{tab:vlaps}).

\subsubsection{Spectral Analysis}

Due to the limited quality of the $1860$~MHz map we were unable to calculate a reliable spectral index for the halo core. \citet{Feretti1997} calculate spectral indices using fluxes from the NCRT at $408$~MHz and WSRT at $610$~MHz and $1400$~MHz. They report a steepening spectrum with frequency: $\alpha^{408}_{610} = 0.92$ and $\alpha^{610}_{1400} = 2.2$. Using our new flux from the VLA of $240$~mJy at $1348$~MHz, the spectral index is reduced to $\alpha^{610}_{1348} = 1.8$. However, the discovery of a signficantly larger emitting region with GBT from \citet{Farnsworth2013} indicates that these interferometric observations are missing a substantial amount of flux ($328$~mJy from the GBT {\it vs} $153$~mJy from the WSRT \citep{Feretti1997} at $1400$~MHz), so this steepening must be viewed as tentative. 

\section{X-ray Analysis}\label{sec:Xray}

A2319 has been observed by several X-ray telescopes, including \textit{ASCA} \citep{Markevitch1996}, ROSAT \citep{Feretti1997}, \textit{BeppoSAX} \citep{Molendi1999}, \textit{Chandra} \citep{Govoni2004a,OHara2004}, \textit{Suzaku} \citep{Sugawara2009}, and \textit{XMM-Newton} (this work). All instruments reveal an asymmetric X-ray distribution, with the brightest emission located near the center of the A2319A main cluster, and a tail extending to the NW towards the A2319B subcluster. It is a relatively hot cluster, with a mean X-ray temperature between $9$-$12$~keV, depending on the instrument. Observations from \textit{ASCA}, ROSAT, and \textit{BeppoSAX} revealed temperature decreases to the NW of the emission peak, suggesting that this cooler temperature is associated with the ICM of A2319B. There is no evidence for nonthermal X-ray emission from observations with \textit{BeppoSAX} \citep{Molendi1999}, \textit{Suzaku} \citep{Sugawara2009}, or \textit{Swift} \citep{Ajello2009}.

Temperature maps of A2319 from \textit{Chandra} observations show evidence of cooler regions in the cores of the merging subclusters, and hotter regions perpendicular to the merger axis, consistent with other observations of merging clusters \citep{Govoni2004a,OHara2004}. \citet{Govoni2004a} find for a sample of clusters with radio halos that in general the radio halo tends to trace the hotter X-ray regions. However, these temperature maps are only sensitive to the central, brightest region of the cluster, so it is difficult to characterize the relationship between the large-scale halo and the X-ray temperature.

A detailed study of the merger history of A2319 using \textit{Chandra} observations is found in \citet{OHara2004}. There is a clear discontinuity seen in the \textit{Chandra} X-ray image $\sim3'$ to the SE of the brightness peak, which is identified as a cold front. The peak X-ray emission is offset from the central cD galaxy. \citet{OHara2004} also find evidence for dimmer emission in the region of A2319B. The authors propose a scenario in which A2319 is post merger, and the two subcluster cores are moving apart. In this scenario, A2319B moved past the main core with a nonzero impact parameter and was stripped of most of its gas, while the core of A2319A was displaced from its pre-merger position. The interaction between the cold core of A2319A and the surrounding warmer ICM is responsible for the formation of the cold front. They argue that these X-ray features, along with information on velocity dispersion from optical analyses, point to a NW-SE merger axis that is $\sim65^{\circ}$ out of the plane of the sky. If this merger is taking place at this large angle to the plane of the sky, then quantitative analyses of this cluster become difficult due to projection effects. 

\subsection{\textit{XMM-Newton} Analysis}

\subsubsection{Data Reduction}

We analyzed the three archival \textit{XMM} observations of A2319 (ObsIDs: 0302150101, 0302150201, 0600040101), using data from the MOS1, MOS2 and PN cameras on the EPIC instrument. We utilized the XMM Extended Source Analysis Software (XMM-ESAS; \citealt{Kuntz2008,Snowden2008}), in conjunction with the XMM Scientific Analysis System (SAS) version 13.5.0, for data preparation and background modeling. We filtered the data for soft proton flares, masked point sources, and generated quiescent particle background images following the standard ESAS analysis. After filtering, the total exposures for each camera, summed over the three observations, were $\sim80$~ks each for MOS1 and MOS2, and $\sim72$~ks for PN. We created an exposure-corrected, background-subtracted, mosaiced image, binned to $3''$ per pixel, in the soft ($0.5$-$2$~keV) X-ray band.

\subsubsection{Image and Residuals}

\begin{figure} 
  \centering
  \includegraphics[scale=0.36]{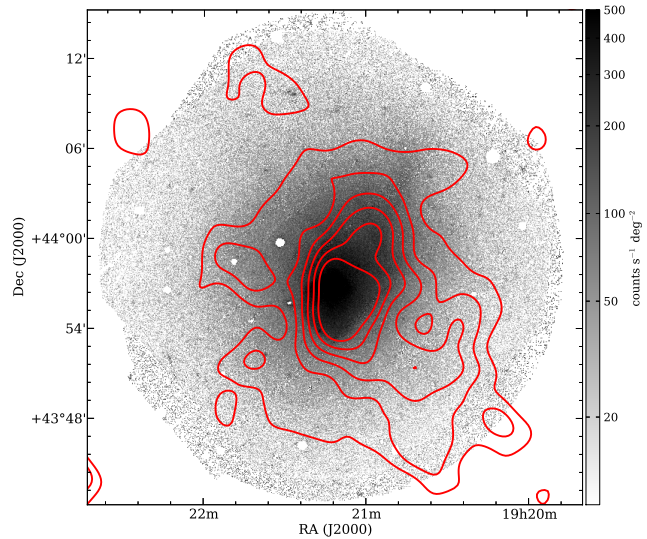}
  \caption{\textit{XMM} observation of A2319, $0.5$-$2$~keV, on a log scale. Pixels are $3''$. $1348$~MHz VLA radio contours are overlaid in red. Levels are (3,6,9...)$\times 0.4$~mJy~beam$^{-1}$.}
  \label{fig:xmm500D0}
\end{figure}

We present an image of the X-ray emission from A2319 in Figure~\ref{fig:xmm500D0}. We clearly observe a surface brightness discontinuity to the SE that is consistent with the previously detected cold front \citep{Govoni2004a,OHara2004,Ghizzardi2010}. A visual inspection of Figure~\ref{fig:xmm500D0} suggests two components to the X-ray emission: a bright core corresponding to the subcluster A2319A, bounded on the SE side by the cold front and extending to the NW in the direction of the subcluster A2319B, and a more symmetric, fainter emission region outside the cold front. There is no obvious sign of excess X-ray emission in the region where the SW extension to the radio halo is found. Our results are consistent with the previous \textit{Chandra} observation. 

Motivated by two-component structure evident in the X-ray emission, we simultaneously fit two smooth, elliptical beta models to the X-ray emission to examine the the residuals \citep{Cavaliere1976,Sarazin1986}. The first beta model is fit to the core region (bounded on the SE by the cold front) and the second is fit to the more symmetric extended emission region:
\begin{align}
  S_{core}(r_{\perp}) &= S_1\left(1+\left(\frac{r_{\perp}}{r_{c1}}\right)^2\right)^{-3\beta_1 +0.5}\\
  S_{ext}(r_{\perp}) &= S_2\left(1+\left(\frac{r_{\perp}}{r_{c2}}\right)^2\right)^{-3\beta_2 +0.5} + S_b
\end{align}
where $S_1$ and $S_2$ are the peak amplitudes in X-ray brightness (in counts~s$^{-1}$~deg$^{-2}$) of each component of the double beta model, $r_{c1}$ and $r_{c2}$ are the two core radii, $r_{\perp}$ is the projected distance from the peak, and $S_b$ is a constant background term. The two beta model fits have slightly different centers.
\begin{figure} 
  \centering
  \includegraphics[scale=0.36]{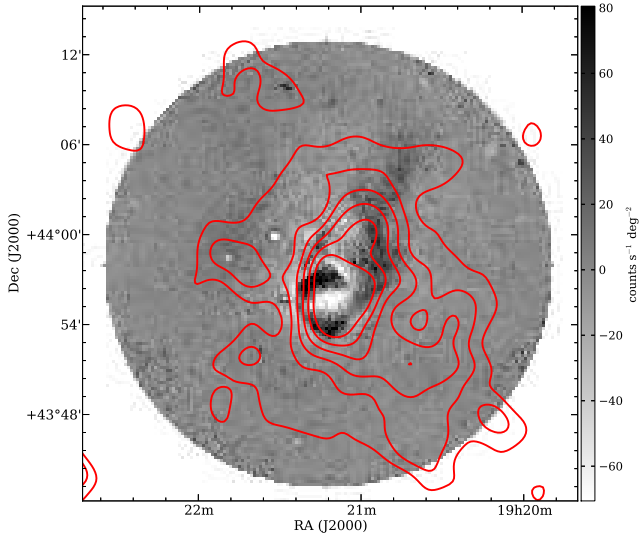}
  \caption{\textit{XMM} residuals, $0.5$-$2$~keV, after subtraction of a double elliptical beta model. Pixels are $12''$. Radio contours from VLA at $1348$~MHz are in red. Levels are (3,6,9...)$\times 0.4$~mJy~beam$^{-1}$.}
  \label{fig:xmmres500}
\end{figure}
We binned the X-ray image to $12''$ per pixel and fit the data using the package \textit{Sherpa}. The reduced chi-squared for our best fit is $2.3$ for $17178$ degrees of freedom.The best fit values for the core radii $r_{c1}$ and $r_{c2}$ are $128$~kpc and $394\pm9$~kpc, respectively. The value for $r_{c1}$ is at its maximum bound (corresponding to the distance from the X-ray peak to the cold front). Best-fit values for $\beta_1$ and $\beta_2$ are $0.644\pm0.005$ and $0.77\pm0.02$, respectively. The best fit background value is $9.4\pm0.2$~counts~s$^{-1}$~deg$^{-2}$. We quote $1\sigma$ statistical uncertainties on best-fit values, but stress that the surface brightness of this cluster is not expected to be well-modelled by any smooth $\beta$-model, given the asymmetry in the X-ray emission due to the cluster merger.

In Figure~\ref{fig:xmmres500}, we see clear evidence of a surface brightness discontinuity to the SE, corresponding to the previously detected cold front. The spiral pattern of positive residuals seen in Figure~\ref{fig:xmmres500} is commonly found in simulations of cluster mergers with nonzero impact parameters, which leaves the more massive core intact and triggers sloshing that produces the cold front \citep{Ricker2001,Ascasibar2006,Roediger2012,Lagana2010}. This interpretation is consistent with the merger picture put forth by \citet{OHara2004}. We do not find any evidence in the residuals for excess emission in the SW region after subtraction of the best fit smooth double beta model.

\section{Discussion}\label{sec:disc}

\subsection{Radio Halo Substructure and the X-ray Cold Front}

In the bright, central region of the cluster, the radio emission traces the X-ray emission remarkably well. The radio brightness at $1348$~MHz falls off rapidly across the cold front, especially visible towards the eastern edge.

In order to examine the profiles of  the X-ray and radio emission across the cold front region, we calculated the average brightness in a 90 degree wedge oriented east-west, and plotted it in Figure~\ref{fig:proj}. Note the distinct change in slope of the X-ray profile across the cold front, steep in the interior (left) and shallower beyond the cold from (right). The same is true for the radio emission, although the transition is significantly broadened because of the $120''$ beam.

\subsection{A Multi-Component Radio Halo}

\begin{figure} 
  \centering
  \includegraphics[scale=0.22]{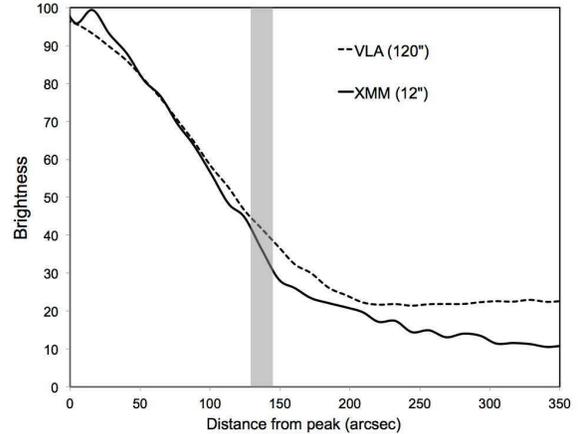}
  \caption{Brightness profiles for X-ray (solid) and radio (dashed), in a 90$^{\circ}$ wedge centered at $\alpha =19h21m0.56s$, $\delta=43^{\circ}56'48''$ and extending east. The X-ray image is averaged over $12''$ annuli. The radio is convolved to a $120''$ beam. The region containing the cold front is highlighted in gray. Brightness is in arbitrary units normalized to the peak.}
  \label{fig:proj}
\end{figure}

The brightness profiles of the core X-ray and VLA radio emission (see Figure~\ref{fig:proj}), together with the substantially larger radio emission detected by the GBT (Figure~\ref{fig:GBT-VLA}), provide evidence for a cluster with distinct emission regions that are perhaps produced by different underlying emission processes. The X-ray emission together with temperature maps from \citet{Govoni2004a} and \citet{OHara2004}, show a distinct, cold X-ray core that has been disturbed by a significant merging event and has compressed some of the ICM near it, producing a cold front. The merger likely occurred with a nonzero impact parameter and at a signficant angle to the plane of the sky. There is additionally a fainter, larger component to the X-ray emission that maps the hotter, more diffuse gas of the ICM; this is possibly gas that was undisturbed by the merger event or has since relaxed.

The radio emission also contains multiple components. There is a large-scale, $2000$~kpc component detected with the GBT. Some of this large scale emission is also seen with the VLA in the SW extension. A smaller, $\sim 800$~kpc brighter region of radio emission is embedded in the larger halo. This radio core closely traces the X-ray emission, and the brightness of this region falls off sharply in the same location as the X-ray cold front.

It is perhaps natural to speculate on the (potentially different) origins for these two components of the radio halo. Cluster mergers drive shocks and tubulence throughout the ICM, providing acceleration sites for the cosmic rays responsible for radio halos \citep[e.g.,][]{Brunetti2014}. The large-scale halo component may be the result of this usual story: merger-generated cosmic ray acceleration that permeates the entire cluster volume. In the hadronic model, long-lived cosmic ray protons continuously resupply the radiating cosmic ray electrons. The lack of X-ray emission in the SW region of the halo implies the lack of thermal electrons, and therefore the lack of cosmic ray proton collision targets. The fact that we observe radio emission in this region already suggests that a hadronic origin in this region is disfavored (see Section~\ref{sec:had}). Cosmic ray proton collisions produce gamma rays in the hadronic model, so limits on the gamma-ray emission from A2319 with \textit{Fermi} could provide even stronger constraints on this model. However, current gamma-ray limits from this cluster and others  already put tension on a hadronic origin for the cosmic rays for $\mu$G magnetic fields \citep[e.g.,][]{Jeltema2011,Brunetti2012b,Ackermann2014}. The alternative for the larger scale component of the halo is that cluster-wide turbulent reacceleration of pre-existing cosmic ray electrons is responsible.

The origins of the smaller radio core, by constrast, may be tied closely to the dynamics of the remaining subcluster core and the X-ray cold front. Simulations of minor mergers (with subcluster mass ratios of approximately 10:1) show that the turbulence generated by core sloshing is confined to the regions inside cold fronts and this turbulence may be responsible for observed radio mini-halos \citep{Zuhone2013}. These mini-halos are typically $<$300~kpc across and are found in a handful of cool-core clusters \citep[e.g.,][]{Feretti2012,Gitti2012}. They are often accompanied by a bright central radio galaxy \citep[e.g.,][]{Blanton2001,Doria2012}.

A2319, by contrast, does not have a cool core; its central entropy of \textbf{K$_0$}$=270$~keV-cm$^2$ \citep{Cavagnolo2009} and subcluster mass ratio of $3:1$ \citep{Oegerle1995} puts it firmly in the recent merger class. Nor does it have a bright central radio galaxy. However, the striking similarity between A2319's radio and X-ray emission raises the question about whether previously suggested models for generating centralized radio halo cores or mini-halo are appropriate.

\subsection{Core Magnetic Field}

Magnetic fields in clusters are essential for discriminating between origin models for radio emission, especially with limited concrete spectral information, but are poorly understood \citep[e.g.,][]{Feretti2012}. We can calculate the volume-averaged magnetic field, $B_{eq}$, from equipartition, by assuming the cosmic ray energy density (protons and electrons) is equal to the magnetic field energy density. We use the revised equipartion formula for $B_{eq}$ derived in \citet{Beck2005}, with the cosmic ray proton to electron number ratio, $k=100$, appropriate for acceleration by either direct, turbulent or secondary, hadronic, processes. This equation does not rely on a choice of integration bounds in frequency space, which, in the classical equipartition calculation, induces an implicit dependence on the magnetic field. 

To calculate the equipartion field, we consider the bright central core of the cluster, limiting the region to that enclosed by the $12\sigma$ contour on the $1348$~MHz radio emission, which also corresponds to the X-ray core (that is, the region inside the cold front). For the line of sight depth of the region, we use $l \sim 500$~kpc, which is approximately the width of the region enclosed by the $12\sigma$ contour. With a brightness of $0.5\mu$Jy~arcsec$^{-2}$ and a spectral index $\alpha = 1.8$, we derive $B_{eq}=2.8~\mu$G. For $\alpha = 0.92$, this decreases to $1.7~\mu$G. These values are consistent with those estimated from Faraday Rotation measures for disturbed clusters \citep[e.g.,][]{Govoni2004}.

\subsection{Comparing the X-ray and Radio: A Test for Hadronic Origins}\label{sec:had}

A spatial comparison of the X-ray and radio emission can help to discriminate between different acceleration models for cosmic rays and probe the structure of magnetic fields in clusters. 

The X-ray emissivity due to thermal bremsstrahlung radiation, the dominant continuum emission mechanism, depends on the thermal ICM density, $n_{th}$ and the X-ray temperature, $T_X$:
\begin{equation}
  \epsilon_X \propto n_{th}^2 T_X^{1/2}
\end{equation}
The temperature in A2319 only changes by a factor of $\la2$ across the cluster \citep{Govoni2004a,OHara2004}; we can therefore safely ignore the weak dependence on temperature. 

The radio emissivity depends on how the cosmic ray electrons responsible for the synchrotron emission are generated. In the case of hadronic origins, assuming a power law distribution for the cosmic ray protons, the synchrotron emissivity depends on the cosmic ray proton density, the thermal ICM density and the magnetic field \citep[e.g.,][]{Brunetti2012b}:
\begin{equation}
  \epsilon_{\nu} \propto \nu^{- \alpha} n_{th}n_{CRp}\frac{B^{1+\alpha}}{B^2+B_{CMB}^2}
\end{equation}
where $\alpha$ is the radio spectral index ($S_{\nu}\propto \nu^{-\alpha}$). Note this expression includes electron injection losses due to synchrotron and Inverse Compton scattering. $B_{CMB}$ is the magnetic field equivalent of the Cosmic Microwave Background energy density, and is equal to $3.25\mu$G at $z=0$. If we assume the cosmic ray proton density $n_{CRp}$ roughly scales with the thermal density $n_{th}$, then the radio emissivity scales with the X-ray emissivity convolved with the magnetic field dependence.

\begin{figure} 
  \centering
  \includegraphics[scale=0.36]{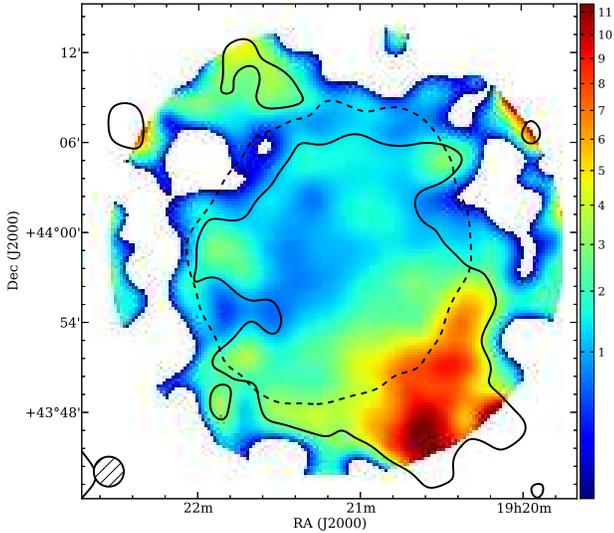}
  \caption{$1348$~MHz radio emission divided by $0.5-2$~keV X-ray emission, arbitrary units. The X-ray image was binned to the same pixel size as the radio image ($12''$) and both the radio and X-ray images were convolved to $120''$ resolution and normalized before dividing. Colors are on a square root scale. $1348$~MHz radio $3\sigma$ contour (solid) and X-ray contour (dashed) at 30 counts~s$^{-1}$~deg$^{-2}$ overlaid in black.}
  \label{fig:XRcomp}
\end{figure}

To make a quantitative comparison between the radio and X-ray images, we convolved the X-ray image to $120''$ resolution and regridded it onto a $12''$ pixel grid. The result of dividing the radio image by the X-ray image, each normalized to a peak of 1, is shown in Figure~\ref{fig:XRcomp}. In the context of a hadronic origin model for the cosmic rays, this yields a spatial map of the magnetic field with an overall scaling dependent on the spectral index:
\begin{equation}
  \frac{\epsilon_{\nu}}{\epsilon_X} \sim \frac{B^{1+\alpha}}{B^2+B_{CMB}^2}
\end{equation}
The ratio between the radio and the X-ray is approximately constant in the central X-ray emitting region. Towards the SW, this ratio grows by a factor of $\ga10$. Assuming hadronic origins for the cosmic rays, this would imply that the magnetic field profile is relatively flat over the central region of the cluster, but increases towards the SW region, in the direction of the new extension to the radio halo. In the case of direct (re-) acceleration by ICM turbulence, the enhanced radio emission would be explained by either increased turbulence or an excess of seed cosmic ray electrons in this region.

There is evidence from simulations and Faraday Rotation measurements of galaxies in clusters that the magnetic field profile should decrease with increasing radius, and roughly follow the thermal electron density \citep{Dolag2001,Govoni2004,Bonafede2010,Donnert2013}. The magnetic field profile inferred from Figure~\ref{fig:XRcomp} is clearly asymmetric, and contradicts this evidence. Alternatively, if the magnetic field is not larger in this SW extension, then the cosmic ray proton density must increase. This is also unlikely, as the cosmic ray proton density is typically assumed to also follow the thermal gas density \citep[e.g.,][]{Pinzke2010}. We therefore argue that a hadronic origin model for the cosmic rays in A2319 is disfavored. At the same time, the turbulent re-acceleration model could be consistent with the data, but there is no way currently to tell whether the requisite enhanced turbulence or seed relativistic electrons are present. Detailed spectral index maps of the radio halo would help to clarify this scenario.

\section{Conclusions}\label{sec:end}

We present results from observations of the merging cluster A2319 with the VLA at $1348$~MHz and \textit{XMM} in the $0.5-2$~keV band. We tentatively report on the discovery of the multi-component nature of the radio halo in A2319: (1) a large-scale, 2~Mpc, component discovered by \citet{Farnsworth2013} with the GBT and partially detected with our VLA observations at $1348$~MHz, and (2) a smaller, 800~kpc radio core that is bounded on one side by a cold front observed in the X-ray. In the X-ray, we confirm the previous detections of the X-ray cold front to the SE and provide strong evidence for core sloshing in the form of a spiral-like structure in the residual X-ray emission after subtraction of a smooth, symmetric component. We also show via a simple spatial comparison of the X-ray and radio emission that a hadronic interpretation for the radio emission, at least outside the X-ray core, is disfavored, due to the lack of X-ray emitting gas (and therefore targets for cosmic ray proton collisions) in that region.

We speculate that these two radio components may have different origins. The large-scale component may be the result of merger-driven turbulence that fills the cluster volume, thus providing acceleration sites for cosmic rays (protons or electrons). The presence of the smaller radio core appears to be related to the motion of the subcluster A2319A core, and could be the result of turbulence related to this core motion that is confined to the cluster core. We propose a scenario in which A2319 recently experienced a significant merger with a nonzero impact parameter that left the more massive cluster core somewhat intact but caused it to slosh around in its gravitational potential well, resulting in a cold front observable in the X-ray and a two-component radio halo. This scenario is consistent with other X-ray studies of this cluster \citep{OHara2004,Govoni2004a}.

A multi-component radio halo is not entirely unprecedented. The cluster A2142, which hosts multiple cold fronts and previously detected radio emission in the cluster center classified as a mini-halo, is now known to also host a giant, $\sim2$~Mpc radio halo \citep{Farnsworth2013} and a fourth cold front $\sim 1$~Mpc from the cluster center \citep{Rossetti2013}. These new discoveries challenge the prevailing paradigm that cleanly separates merging systems with disturbed X-ray emission and giant radio halos from relaxed systems, with cool cores, regular X-ray emission, and mini-halos. The recent discovery of a giant, $\sim 1.1$~Mpc radio halo in the cool-core cluster CL1821+643 \citep{Bonafede2014} further suggests that our current understanding of how mergers and the resulting cluster dynamics impact the production of radio emission needs revision. We add A2319 to this new ambiguous class of clusters that are perhaps in various intermediate stages between relaxed and disturbed systems, leading to novel radio and X-ray morphologies.

Observations of clusters with the next generation of radio instruments, such as LOFAR, ASKAP, and Apertif in the near future, and SKA further out, should provide significant clarity to the increasingly complex picture of how cluster dynamics, and in particular, off-axis merger events, impact cosmic rays, magnetic fields, and the resulting radio emission. Combining observations from the as-yet-unexplored low-frequency ($\sim 10-200$~MHz) band with LOFAR and the increased sensitivity of ASKAP and Apertif at $1.4$~GHz will provide detailed spatial and spectral information that has strong discriminating power between cosmic ray origin models for the origins of cluster radio emission. A detailed spectral index map for A2319 in particular would help to further distinguish the smaller radio core from the larger emission region.

This study of A2319 highlights the need to combine different wavelengths of the same object in order to fully understand the interactions between the thermal and non-thermal components of clusters. In light of this work, we plan a future study that expands on our current analysis of A2319 to include more information on the thermal component with SZ data and the nonthermal component with gamma-ray upper limits.

\section*{Acknowledgements}
We thank Stefano Profumo and Elke Roediger for useful discussions. Partial support for this work at the University of Minnesota comes from grant AST-112595 from the National Science Foundation. E.S. acknowledges support from the Cota-Robles Fellowship. T.E.J. acknowledges support from the Hellman Fellows Fund. The National Radio Astronomy Observatory is a facility of the National Science Foundation operated under cooperative agreement by Associated Universities, Inc. This work is partly based on observations obtained with \textit{XMM-Newton}, an ESA science mission with instruments and contributions directly funded by ESA Member States and the USA (NASA). The X-ray data were provided through the HEASARC \textit{XMM-Newton} archive at NASA/GSFC. This research made use of the NASA/IPAC Extragalactic Database (NED) which is operated by the Jet Propulsion Laboratory, California Institute of Technology, under contract with NASA. This research made use of the software package \textit{Sherpa}, provided by the Chandra X-ray Center. This research made use of Astropy, a community-developed core Python package for Astronomy \citep{Robitaille2013}.

\bibliography{A2319}

\clearpage

\end{document}